\documentclass[11pt]{article}
\pdfoutput=1
\begin{document}
\title{Vortex-filament interactions}
\author{Silas Alben \\
\\\vspace{6pt} School of Mathematics, \\ Georgia Institute of Technology, Atlanta, GA 30032, USA}
\maketitle
\begin{abstract}
We present a fluid dynamics video showing how a point vortex interacts with a passive flexible filament,
for various values of filament bending stiffness.
\end{abstract}
\section{Introduction}
The interaction between solid bodies and vortices is important in many areas of
biology, physics, and engineering. In this work we study a fundamental interaction between a vortex and a
simple passive body---a flexible filament. We consider a point vortex in a purely
inviscid flow. The filament has uniform bending stiffness (with values given in the video frames),
is clamped at its left end, and is free at its right end.

The first sequence shows the motion of a point vortex around a nearly rigid filament. The green circle is centered at the location of the point vortex and the filament is the black line. The trajectory,
shown by the light green line, is nearly periodic. In subsequent clips, we decrease the filament
bending stiffness and show the dynamics. In general, the filament is attracted to the vortex, leading
to a finite-time collision, with distance tending to zero as the square root of temporal displacement from
the collision time.
For some values of bending stiffness, we show dynamics for
different initial positions. In some clips, we also show a second view at the bottom, 
which zooms in on the region of the filament near the vortex.

Our computational method is similar to that in ``Simulating the dynamics of flexible bodies and vortex sheets'' by S. Alben, {\it Journal of Computational Physics}, 228, 2587-2603 (2009).
\end{document}